\documentstyle[prb,epsfig,aps,multicol]{revtex}
\begin{document}
\title{Evolution of the Potential Energy Surface with Size for Lennard-Jones Clusters}
\author{Jonathan P.~K.~Doye, Mark A.~Miller and David J.~Wales}
\address{University Chemical Laboratory, Lensfield Road, Cambridge CB2 1EW, UK}
\date{\today}
\maketitle
\begin{abstract}
Disconnectivity graphs are used to characterize the potential energy surfaces
of Lennard-Jones clusters containing 13, 19, 31, 38, 55 and 75 atoms.
This set includes members which exhibit either one or two `funnels'
whose low-energy regions may be dominated by a single deep minimum or
contain a number of competing structures.
The graphs evolve in size due to these specific size effects and 
an exponential increase in the number of local minima with the number of atoms.
To combat the vast number of minima we investigate the use of 
monotonic sequence basins as the fundamental topographical unit.
Finally, we examine disconnectivity graphs for a transformed energy
landscape to explain why the transformation provides a useful approach
to the global optimization problem.
\end{abstract}

\begin{multicols}{2}
\section{Introduction}

The relationship between dynamics and the potential energy surface (PES), 
or potential energy `landscape', 
holds the key to understanding a wide range of phenomena, 
including protein folding, global optimization
and the properties of glasses. 
For example, what are the topographical features of the PES that enable 
a protein to fold to its native state? 
What sort of PES renders global optimization tractable,
and how can it be transformed\cite{StillW88} to make global optimization easier?
Which features of the PES produce good glass formers, and what distinguishes
`strong' from `fragile' liquids\cite{Angell95}?

Stillinger and Weber\cite{StillW84a} first
showed the utility of partitioning configuration
space into basins of attraction surrounding each local minimum. The thermodynamics of 
the system can then be formulated in terms of the properties of these local minima.\cite{Wales93a,Doye95a}
However, the minima alone contain no dynamic information.
Therefore, to obtain insight into the dynamics it is necessary to locate 
the transition states which connect the local minima. 
This information can then be incorporated into a master equation that describes 
the flow of probability between minima as the system evolves towards 
equilibrium.\cite{Czerminski90,Kunz95,Angelani98} 
As well as providing a picture of the relationship between the dynamics and the PES,
this approach can probe time scales much longer than those accessible in
conventional simulations.\cite{MarkPhD} 

We also wish to relate the thermodynamics and dynamics to general features
of the PES in a qualitative manner.
Disconnectivity graphs are a recently-developed tool that provide
a helpful visualization of the PES. These graphs have now been calculated for a number
of polypeptides\cite{Czerminski90,Becker97,Levy98a} and various clusters.\cite{WalesMW98,Miller99a,Doye99c}
They show which of the minima in a sample are connected by pathways lying below 
a series of energy thresholds, and so provide a picture of the hierarchy of energy barriers 
in a system.

In this paper we present disconnectivity graphs for a series of Lennard-Jones (LJ) clusters.
Our aims are to gain more insight into the potential energy landscapes of these clusters
and to explore systematically how the graphs evolve with size.
LJ clusters provide an ideal test case because their dynamics and thermodynamics
have been thoroughly studied and they have been used in the
development of global optimization algorithms.\cite{Pillardy,Leary97,WalesD97} 

As the size of the cluster increases the number of minima and 
transition states is expected to grow at least exponentially, 
even when permutational isomers are not included.\cite{HoareM76,Tsai93a,Still99}
Furthermore, cluster properties do not vary smoothly with the number of atoms in this size regime,
and we have chosen to examine sizes which should illustrate particularly interesting features.

Due to the exponential increase in the number of stationary points with size
it is difficult for a disconnectivity graph to retain a global picture of the PES.
We have therefore explored the use of monotonic sequence basins\cite{BerryK95}
as the fundamental topographical unit rather than minima (Section \ref{sect:mono}).  
Finally, in Section \ref{sect:gmin} we use disconnectivity graphs to probe the transformed energy 
landscape that is searched by the Monte Carlo minimization\cite{Li87a} 
or basin-hopping\cite{WalesD97} algorithm. 

\section{Methods}
\label{sect:meth}

The atoms interact via a Lennard-Jones potential\cite{LJ}:
\begin{equation}
E_c = 4\epsilon \sum_{i<j}\left[ \left(\sigma\over r_{ij}\right)^{12} - \left
(\sigma\over r_{ij}\right)^{6}\right],
\end{equation}
where $\epsilon$ is the pair well depth and $2^{1/6}\sigma$ is the
equilibrium pair separation.

To examine the topography of the PES, we need to locate its minima and
the network of transition states and pathways that connect them.
Transition states can be found efficiently using 
eigenvector-following,\cite{Pancir74a,Cerjan,Wales94b} in which
the energy is maximized along one direction and simultaneously minimized in all the others. 
The minima connected to a transition state are defined by 
the two steepest-descent paths which begin
parallel and antiparallel to the unique Hessian eigenvector whose corresponding eigenvalue
is negative. 
The steepest-descent paths were calculated using a method which employs analytic second
derivatives.\cite{Page88}

The samples of minima and transition states were generated using an approach 
which has also been described before.\cite{Miller99a,Doye99c,Tsai93a}
The elementary step in this process is to perform a transition
state search after stepping away from a minimum along one of the Hessian eigenvectors
and then, if this search finds a transition state, to generate the corresponding pathway.
If the transition state is connected to the 
minimum from which the search was started, we add it to our database along with
the connected minimum.
Occasionally we find that the transition state is not connected to the starting minimum.
In this case we only add it to the database if it is connected to 
a minimum that is already present.
At each step we start from the lowest-energy minimum from which fewer
than a specified number, 2$n_{\rm ev}$, of transition state searches have been performed. 
Searches are carried out in positive and negative directions
along each eigenvector in order of increasing eigenvalue.
The value chosen for $n_{\rm ev}$ determines how thoroughly the PES is searched
in the vicinity of a given minimum.

By repeating this process until no new minima are found
we can obtain a nearly exhaustive catalogue of the minima.
This approach was used for the 13-atom cluster, LJ$_{13}$, but
for the larger clusters it is impossible to find all the minima. 
We therefore stopped searching once we were confident
that we had obtained an accurate representation of the low-energy regions of the PES. 
For LJ$_{75}$ an alternative strategy was required due to the double-funnel character
of the landscape; this approach is described in the following section.

\begin{center}
\begin{figure}
\epsfig{figure=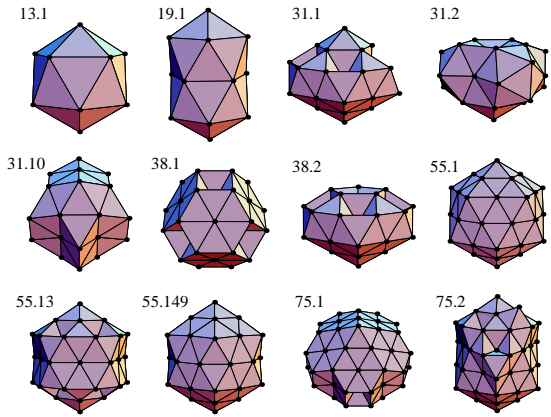,width=8.2cm}
\vglue-0.3cm
\begin{minipage}{8.5cm}
\caption{\label{fig:structures} 
A selection of low-energy minima for LJ$_{13}$, LJ$_{19}$, LJ$_{31}$, LJ$_{38}$, LJ$_{55}$ and LJ$_{75}$.
The first number denotes the number of atoms, 
and the second denotes the energetic rank of the minimum
(the global minimum has rank 1, etc.).
}
\end{minipage}
\end{figure}
\end{center}

\section{Results}

\begin{center}
\begin{figure}
\epsfig{figure=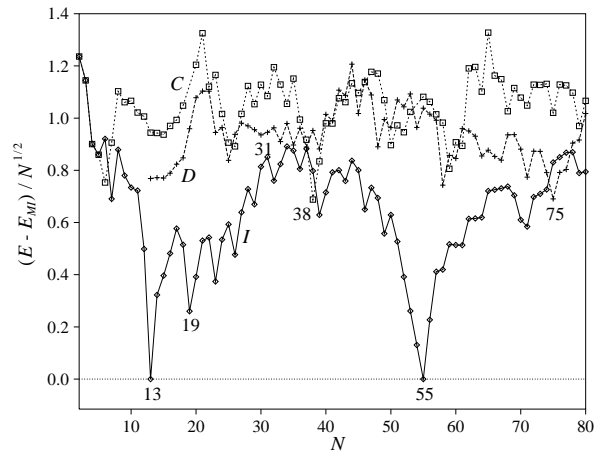,width=8.2cm}
\vglue0.1cm
\begin{minipage}{8.5cm}
\caption{\label{fig:EvN} 
Comparison of the energies of icosahedral (I), decahedral (D) and 
close-packed (C) LJ$_N$ clusters. 
The energy zero is $E_{\rm MI}$, a function fitted to the energies of 
the first four Mackay icosahedra at $N$=13, 55, 147 and 309:
$E_{\rm MI}=-2.3476-5.4633N^{1/3}+14.8814N^{2/3} - 8.5699N$.
The sizes that we consider in detail in this paper have been labelled.
}
\end{minipage}
\end{figure}
\end{center}

We now characterize the PES's of 
LJ$_{13}$, LJ$_{19}$, LJ$_{31}$, LJ$_{38}$, LJ$_{55}$ and LJ$_{75}$.
To explain this selection of cluster sizes, and as a background to the interpretation
of the disconnectivity graphs, we briefly review some of the structural properties 
of small LJ clusters. Comparisons of particularly stable sequences of clusters indicate that 
structures based on Mackay icosahedra\cite{Mackay} 
(e.g.\ 13.1 and 55.1 in Figure \ref{fig:structures}) are 
dominant up to $N\sim 1600$.\cite{Raoult89a} Therefore, as can be seen from Figure \ref{fig:EvN},
most of the clusters in the size range that we are considering have a global minimum 
based on icosahedral packing.
The sizes at which complete Mackay icosahedra are geometrically
possible ($N$=13, 55,$\ldots$) are 
particularly stable compared to other sizes,
leading to particularly low-energy global minima (Figure \ref{fig:EvN}).
For these clusters, there is a large energy gap between the two lowest-energy minima 
($2.85\epsilon$ and $2.64\epsilon$ for LJ$_{13}$ and LJ$_{55}$, respectively)
and we expect the PES to have a single deep funnel. 

The term `funnel' was introduced
in the protein folding literature to describe the situation where a collection of downhill 
pathways converge on the native structure of the protein.\cite{Leopold,Bryngelson95}
As a result of this PES topography
the protein is `funnelled' towards the native structure on relaxation.
We use the term here in a similar manner except that, 
instead of converging on the native structure of the protein, 
the funnel should converge on a single low-energy minimum,
or possibly to a collection of structurally-related low-energy minima.

Between $N$=13 and 55 
the energy first increases relative to an energy function interpolated between the Mackay icosahedra
as an overlayer is added to the 13-atom icosahedron, and then decreases as the overlayer approaches
completion at $N$=55 (Figure \ref{fig:EvN}). 
Of course, there are smaller features superimposed on this broad maximum.
At certain sizes particularly stable overlayers with no low-coordinate atoms can be formed, 
giving rise to subsidiary minima in Figure \ref{fig:EvN}.
For example, at $N$=19 a six-atom cap on the 13-atom icosahedron leads to the relatively
stable double icosahedron shown in Figure \ref{fig:structures}.
We would expect LJ$_{19}$ to exhibit a single-funnel PES but to relax less efficiently
to the global minimum than LJ$_{13}$ or LJ$_{55}$.

We also chose to study a size, $N$=31, from near the top of the broad maximum in Figure \ref{fig:EvN},
where there are a large number of low-energy minima 
associated with the various ways that the atoms of the overlayer on the 13-atom
icosahedron can be arranged. Moreover, for this size the lowest-energy decahedral cluster is only 
a little higher in energy than the icosahedral structures. 

For most sizes the global minimum has icosahedral character, but 
there are a few clusters for which this is not the case.
For $N$=38 the global minimum is the face-centred-cubic (fcc) 
truncated octahedron\cite{Pillardy,Doye95c}
(38.1 in Figure \ref{fig:structures}) and 
for $75\le N\le 77$
the global minima are based on the Marks decahedron\cite{Doye95c} 
(75.1 in Figure \ref{fig:structures}).
We chose to study $N$=38 and 75 where the global minimum is structurally very 
different from the second lowest-energy minimum, which in each case is icosahedral.
The two lowest-energy minima are therefore rather distant in
configuration space, and they are separated by large potential energy\cite{Doye98e} 
and free energy\cite{Doye99c} barriers.
Therefore, we expect these surfaces to have two funnels: one which leads 
to the low-energy icosahedral minima and one which leads to the global minimum.

Details of the samples of minima and the transition states that we found
for each of the clusters are given in Table \ref{table:nsp}.
For LJ$_{13}$, we were able to gauge how close the samples were to convergence
by following the number of minima and transition states as $n_{\rm ev}$ was increased. 
The number of minima seemed close to reaching an asymptote, but the number 
of transition states was still increasing at $n_{\rm ev}$=15.\cite{MarkPhD}

\subsection{Disconnectivity graphs}

The conceptual basis for the disconnectivity graph approach is as follows.\cite{Becker97}
At a given total energy, $E$, minima can be grouped into disjoint sets,
called superbasins, whose members are mutually accessible at that energy.
In other words, each pair of minima in a superbasin are connected directly or
through other minima by a path whose energy never exceeds $E$,
but would require more energy to reach a minimum in another superbasin.
At low energy there is just one superbasin---that containing the global minimum.
At successively higher energies, more superbasins come into play as new
minima are reached. At still higher energies, the superbasins coalesce
as higher barriers are overcome, until finally there is just one containing
all the minima (provided there are no infinite barriers).

\begin{center}
\begin{table}
\begin{minipage}{8.5cm}
\caption{\label{table:nsp}
The number of minima $n_{\rm min}$, transition states $n_{\rm ts}$, 
and monotonic sequence basins $n_{\rm MSB}$ in our samples.
For each sample, transition state searches were performed from the
lowest $n_{\rm search}$ minima, and for these minima transition state searches
were performed along the lowest $n_{\rm ev}$ eigenvalues. 
Values of $n_{\rm MSB}$ and  $n_{\rm search}$
are not included for LJ$_{75}$ because a different method was used to search the PES.
}
\begin{tabular}{cccccc}
 $N$ & $n_{\rm min}$ & $n_{\rm ts}$ & $n_{\rm ev}$ & $n_{\rm MSB}$ & $n_{\rm search}$ \\
\hline
13 & 1467 & 12435 & 15 & 1 & 1467 \\
19 & 3000 & 9847 & 10 & 22 & 727 \\
31 & 6000 & 9320 & 10 & 124 & 1070 \\
38 & 6000 & 8633 & 10 & 116 & 1271 \\
55 & 6000 & 13125 & 10 & 56 & 2464 \\
75 & 17581 & 21371 & 3 & - & - \\
\end{tabular}
\end{minipage}
\end{table}
\end{center}

To construct a disconnectivity graph, the superbasin analysis is performed at a series
of energies. At each energy, a
superbasin is represented by a node, and lines join nodes in one level to
their daughter nodes in the level below. The horizontal position of a node
has no significance, and is chosen for clarity. Every line terminates at a
local minimum.

\begin{center}
\begin{figure}
\epsfig{figure=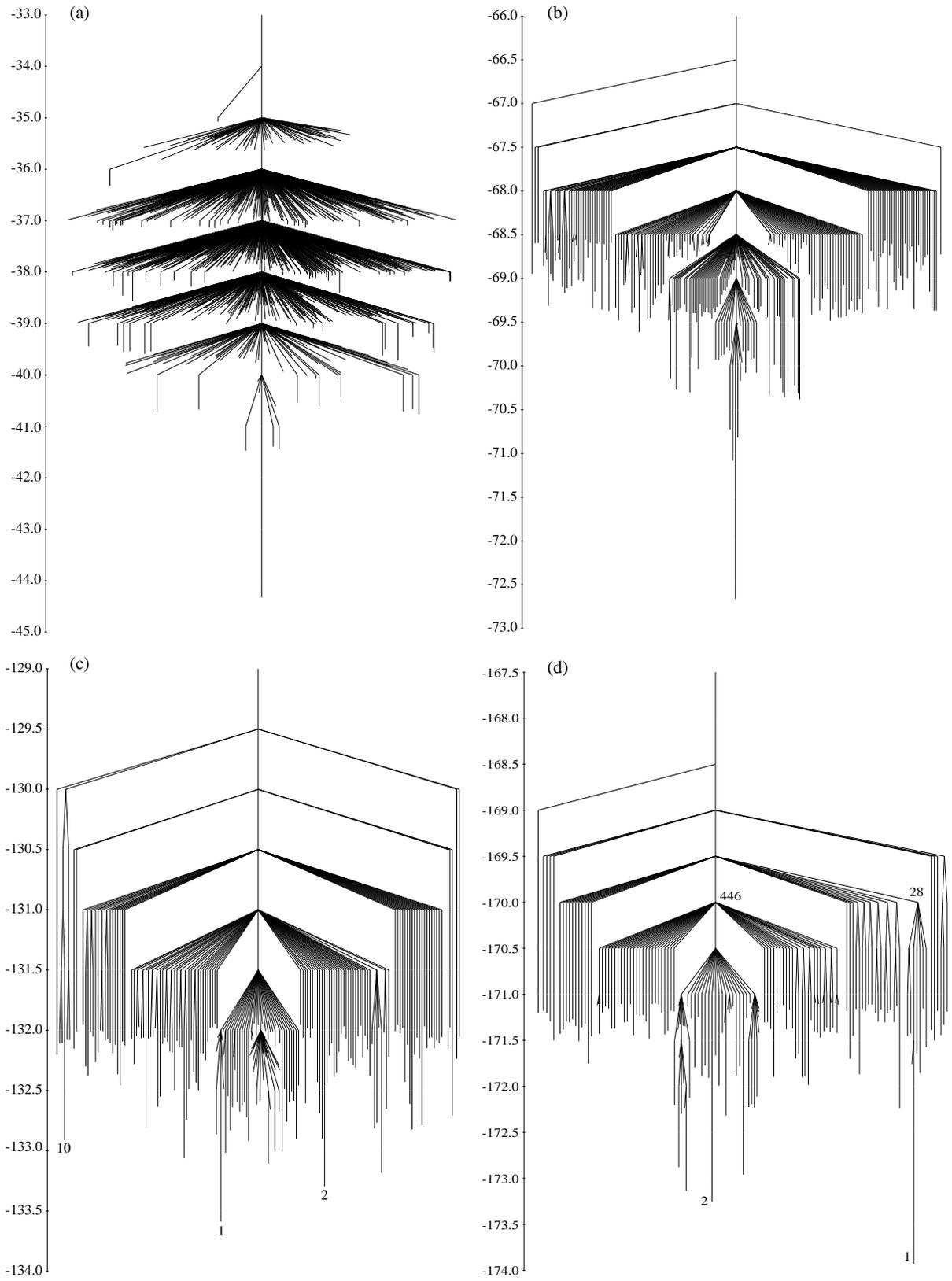,width=16.0cm}
\vglue0.1cm
\begin{minipage}{17.4cm}
\caption{\label{fig:tree} Disconnectivity graphs for 
(a) LJ$_{13}$, (b) LJ$_{19}$, (c) LJ$_{31}$, 
(d) LJ$_{38}$, (e) LJ$_{55}$ and (f) LJ$_{75}$.  
In (a) all the minima are represented. 
In the rest only the branches leading to the 
(b) 250, (c) 200, (d) 150, (e) 900 and (f) 250 lowest-energy minima are shown. 
The numbers adjacent to the nodes indicate the number of minima the nodes represent.
The branches associated with the minima depicted in Figure \ref{fig:structures} 
are labelled by their energetic rank. In (e) an enlarged view of the branch marked i is
shown in the inset. The energy scale is in units of $\epsilon$.
}
\end{minipage}
\end{figure}
\end{center}

The disconnectivity graphs produced using this procedure 
are shown in Figure \ref{fig:tree}.
For $N$=13 it is possible to include all the minima that we have found in the disconnectivity graph,
which therefore provides a practically complete global picture of the PES. 
The graph has the form expected for an ideal funnel: 
there is a single stem, representing the superbasin of the global minimum, with 
branches sprouting directly from it at each level, indicating the progressive 
exclusion of minima as the energy is decreased. 

The form of this graph implies
that all the minima are directly connected to the superbasin associated with the global minimum.
In fact, 99\% of the minima are within two rearrangements of the global minimum
and none are further than three steps away.
Furthermore, there are 911 structurally 
distinct transition states connecting 535 minima directly to the global minimum.
If the reaction path degeneracy is taken into account,
there are $108\,967$ transition states (some of which are permutational isomers) connected to 
each permutational isomer of the global minimum.\cite{sdvef} 

If we examine the distance of a minimum from the
global minimum as a function of its potential energy, 
for LJ$_{13}$ the resulting plot (Figure \ref{fig:S}a)
has the form we would expect of a funnel: the distance from the global minimum 
increases as the potential energy of the minimum increases. 
It is also worth noting that the slope of 
the funnel is steeper than for any of the other clusters.
Hence it is no surprise that relaxation down the PES to the global 
minimum is relatively easy for this cluster. 
However, relaxation is hindered somewhat by the fairly large
barriers (1--2$\epsilon$) that exist for escaping from some of the minima into the superbasin of 
the global minimum (Figure \ref{fig:tree}a). 

\end{multicols}

\addtocounter{figure}{-1}
\begin{center}
\begin{figure}
\epsfig{figure=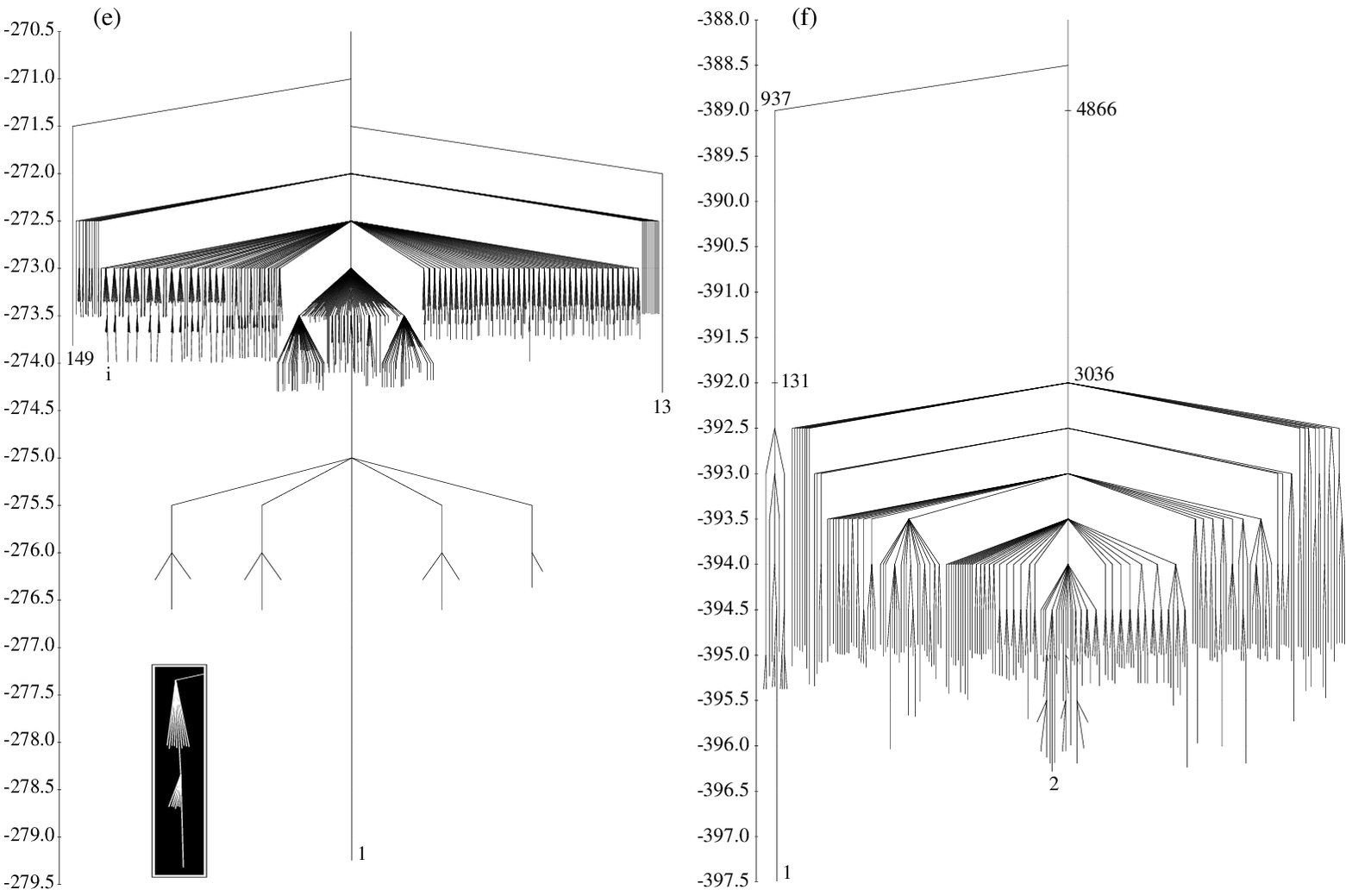,width=16.5cm}
\vglue0.1cm
\begin{minipage}{17.6cm}
\caption{cont}
\end{minipage}
\end{figure}
\end{center}

\begin{multicols}{2}

For all the other clusters it is not computationally feasible to obtain a nearly complete
set of minima. Moreover, if we attempt to represent all the minima of our samples on a 
disconnectivity graph, the density of lines simply becomes too great.
Instead we only represent those branches that lead
to a specified number of the lowest-energy minima, both for clarity and because our samples of minima 
and transition states are likely to be most complete for the low-energy regions of the PES. 
We should note that the minima that are not represented can still contribute to the appearance of the graph
if they mediate low barrier paths between minima that are included. 
However, this approach does have the consequence that, as the size increases, 
the graphs increasingly focus on a smaller and smaller proportion of the surface. 

We can see this effect if we examine the graph
of LJ$_{55}$ for which we also expect the PES to exhibit a single funnel.
Although there is more fine structure,
the form is similar to that of LJ$_{13}$. 
However, all the minima represented in the disconnectivity graph have relatively ordered structures
and so, unlike the graph for LJ$_{13}$, this representation
does not tell us whether there is a funnel leading 
from the disordered liquid-like minima to the global minimum Mackay icosahedron.
Instead, it only shows that the low-energy region of the PES associated with structures based on
the Mackay icosahedron is funnel-like. The graph probably only represents 
the bottom of a larger funnel leading down from the huge number of disordered minima.
The plot of the distance from the global minimum as a function of potential
energy shows a glimpse of this larger funnel (Figure \ref{fig:S}e).

It is easy to understand the differences between the graphs for LJ$_{13}$ and LJ$_{55}$.
The energy gap between the global minimum and the disordered liquid-like minima
is an extensive quantity, whereas the energy gap
between two ordered structures, where a few atoms have changed position, is 
related to the change in the number of nearest-neighbour contacts and
is independent of size. Furthermore, the number of possible ways of arranging
surface atoms and vacancies in an ordered structure becomes increasingly large. 
Therefore, the number of ordered minima which lie below the liquid-like band of minima
increases rapidly with the number of atoms. 

The fine structure of the LJ$_{55}$ disconnectivity graph reveals some interesting features.
The minima separate into bands related to the number of defects present 
in the Mackay icosahedron.\cite{Doye95b}
For example, all the minima in the first band above the global minimum
are Mackay icosahedra with a missing vertex and an atom on the surface. 
The eleven minima in this band correspond to the eleven possible sites 
for this atom that are unrelated by symmetry. Of these minima 
the four lowest in energy have the atom located over the centre of 
one of the faces of the Mackay icosahedron, and the other seven have the atom off-centre.
In the disconnectivity graph these minima split into four groups corresponding
to the four symmetry-unrelated faces on which the adatom can be located.
The splitting occurs because the barriers for rearrangements in which the adatom passes 
between faces are larger 
than the barriers for changing the position of the adatom on a face. 
Therefore, in the disconnectivity graph 
the minima first become connected to the other (one or two) minima with the adatom on the same face
before they are connected to the stem associated with the global minimum.

The second band of minima consists of Mackay icosahedra with two missing vertices and
two surface atoms. The lower-lying minima in this band have the two adatoms in contact,
either on the same face or bridging an edge.
The minima with the two adatoms unpaired give rise to a repeated motif,
an example of which is illustrated in the inset.
This feature is  repeated over ten times on the left-hand side of the graph with the top node 
always at $-273.0\epsilon$.
The minima split into these sets because of the lower barriers for an 
adatom moving between sites on the same face. 
If the two faces with adatoms are unrelated by symmetry, 
there are sixteen distinct ways of arranging the atoms on the two faces. 
The lowest-energy minimum of these sixteen has the two adatoms in the central
sites. Slightly higher in energy are the six minima with one adatom central and one off-centre. 
Higher still are the nine minima with both adatoms off-centre. 

The features noted above are reflected in the dynamics of LJ$_{55}$.
Just below the melting point the cluster passes between the Mackay icosahedron
and states with one and two defects, residing in each state for periods of the order
of nanoseconds (using potential parameters appropriate to argon).\cite{Kunz93,Kunz94} 
While in one of these states, the surface atoms and vacancies diffuse across the surface.
This time scale separation between diffusive motion and the formation and annihilation 
of defects is a consequence of the higher barriers for the latter processes and the distances
over which a surface atom and vacancy must diffuse to recombine. 
The latter is reflected in the wide range of distances that minima
in this second band are away from the global minima.

The two minima (55.13 and 55.149) in Figure \ref{fig:tree}e 
that must overcome the highest barriers to reach the superbasin 
containing the global minimum do not have icosahedral structures.
Both have $D_{5h}$ point group symmetry and are depicted in Figure \ref{fig:structures}.
The lower-energy of the two is actually the thirteenth lowest-energy minimum
and was recently discovered by Wolf and Landman.\cite{Wolf98} 
The other minimum has previously been noted by Wales.\cite{Wales93c}
Both can be converted into a Mackay icosahedron by a single cooperative 
rearrangement in which parts of the structure twist around the fivefold axis.\cite{Wales93c}
As a significant number of nearest-neighbour contacts are broken in these rearrangements,
the barriers are higher than those for the localized rearrangements by which the defective
Mackay icosahedra interconvert.

The disconnectivity graph for LJ$_{19}$ shows that the PES is again funnel-like
(Figure \ref{fig:tree}b), as expected.
The branches for most of the minima connect directly to the superbasin 
containing the global minimum. 
Although our graph only shows branches leading to the lowest 250 minima, 
Figure \ref{fig:S}b reveals that the funnel continues up to higher energies.

The PES of LJ$_{19}$ has been analysed previously in several studies.\cite{Kunz95,BerryK95,Ball96} 
The profiles of the downhill pathways to the global minimum were described 
as `sawtooth-like' rather than `staircase-like' 
because the barrier heights are relatively large compared to the energy differences between the minima. 
On this basis Ball {\it et al.}\/ concluded that LJ$_{19}$ has 
topographical features typical of a glass-former. 
The disconnectivity graph also shows that some of the downhill barriers 
are quite large. However, as the global minimum is at the bottom of a funnel
the barriers only slow down the rate of relaxation towards the double 
icosahedron, rather than preventing it. 
Lowering the energy does not take the system away from the global minimum, 
but rather towards it (Figure \ref{fig:S}b).
The appearance of $N=19$ as a magic number in mass spectra of rare gas clusters\cite{Echt81}
confirms that the global minimum is kinetically accessible.

The disconnectivity graph for LJ$_{31}$ (Figure \ref{fig:tree}c) is fundamentally different from 
those we have considered so far. The energetic bias towards the global minimum 
is smaller (Figure \ref{fig:S}a). In fact there are a number of minima
with energies close to that of the global minimum which are separated from it
by fairly large energy barriers. This situation is partly the result of competition 
between two distinct types of overlayer.
In the first type, the anti-Mackay overlayer, atoms are added to the faces and vertices
of the underlying 13-atom icosahedron (giving
rise, for example, to the double icosahedron, 19.1). In the second type, the Mackay overlayer
atoms are added to the edges and vertices. The completion of the Mackay overlayer 
leads to the next Mackay icosahedron. LJ$_{31}$ is the first size for which 
a cluster with the Mackay overlayer is the global minimum.\cite{Northby87} 
It can be seen from Figure \ref{fig:structures} that minimum 31.1 is a fragment
of the 55-atom Mackay icosahedron. The second lowest-energy minimum, 31.2, 
has an anti-Mackay overlayer. 

There are also some low-energy decahedral minima for LJ$_{31}$. 
For example, minimum 31.9 (Figure \ref{fig:structures})
is separated by a large energy barrier from the global minimum, because
not only must there be a change in morphology from decahedral to icosahedral,
but the cluster must also change shape. 
Some of the decahedra with more spherical shapes
are connected to the superbasin associated with the global minimum by smaller barriers.

We can deduce something of the relaxation behaviour of LJ$_{31}$ from its
disconnectivity graph. Once the cluster has reached a low-energy 
configuration, presumably by rapid descent of a funnel from the liquid,
subsequent relaxation towards the global minimum may be considerably slower.
There is little energetic bias at the bottom of the PES to guide the system towards
the global minimum and the barriers for interconversion of the low-energy
minima can be relatively large. Therefore, it is not surprising that the time
required to find the global minimum using the basin-hopping global optimization 
algorithm shows a maximum at $N$=31.\cite{WalesS99}
 
\end{multicols}

\begin{center}
\begin{figure}
\epsfig{figure=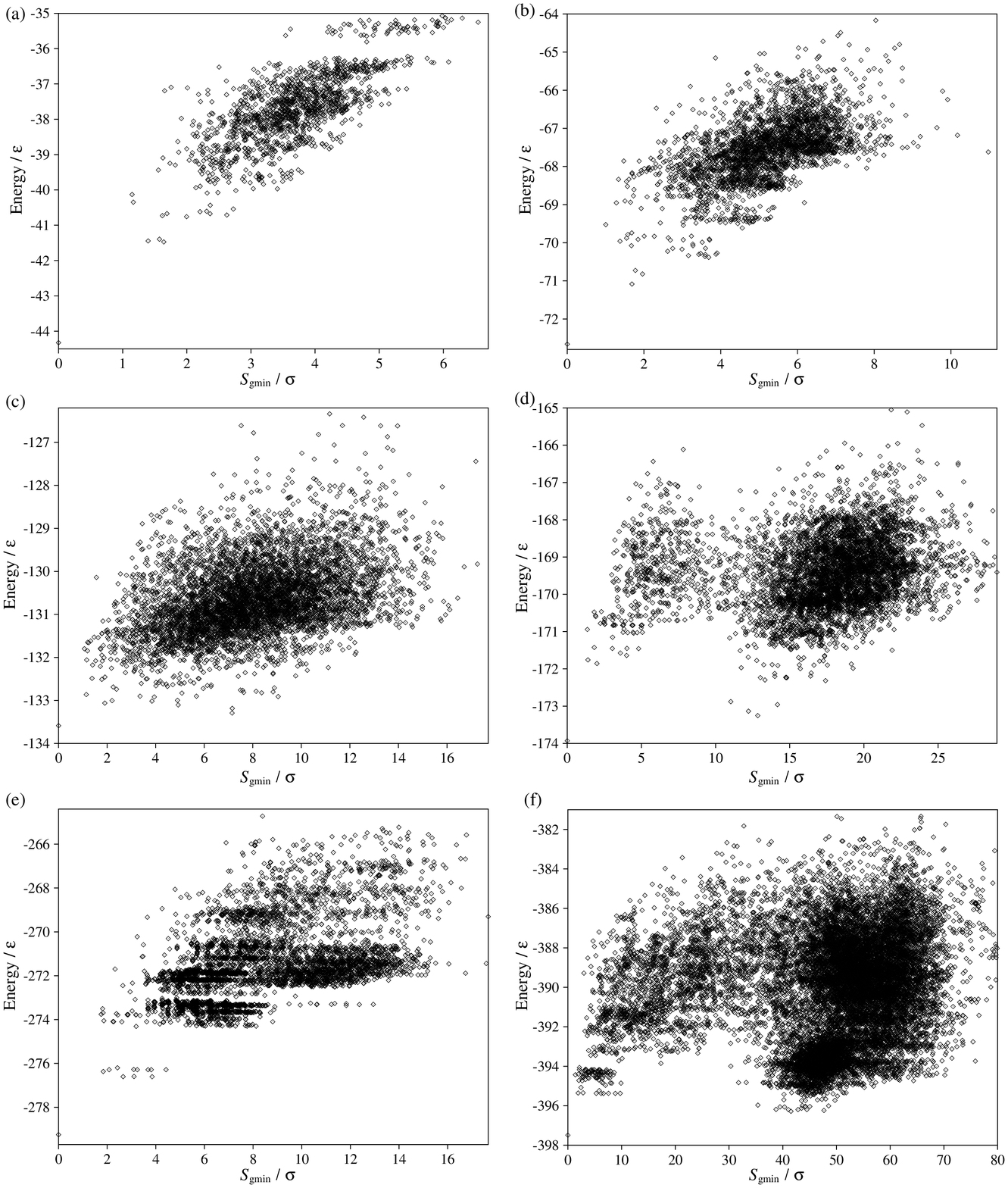,width=16.0cm}
\vglue0.1cm
\begin{minipage}{17.6cm}
\caption{\label{fig:S} 
The dependence of $S_{\rm gmin}$, the integrated path length of the shortest pathway
to the global minimum, on potential energy for our samples of 
(a) LJ$_{13}$, (b) LJ$_{19}$, (c) LJ$_{31}$, (d) LJ$_{38}$, (e) LJ$_{55}$ and (f) LJ$_{75}$ minima.
}
\end{minipage}
\end{figure}
\end{center}

\begin{multicols}{2}

The effects of competing structures that we noted for LJ$_{31}$ appear in a more extreme 
form for LJ$_{38}$ and LJ$_{75}$.\cite{38again} 
For both these clusters the disconnectivity graph clearly separates 
the low-energy minima into two main groups, namely those associated with the global minimum 
and those with icosahedral structure.
These two groups of minima are separated by a large energy barrier,
so the graph splits into two stems at high energy which lead  down to two 
structurally distinct sets of low-energy minima.
This splitting is characteristic of a multiple funnel PES. 
The separation is particularly dramatic for LJ$_{75}$, where the 
decahedral to icosahedral barrier is over $3\epsilon$ larger than any of 
the other barriers between the 250 lowest-energy minima.
Although the disconnectivity graphs clearly show only the bottom of the PES,
these two groups of minima can be associated with separate funnels,
and give rise to distinct thermodynamic states.\cite{Doye99c}
The double funnel structure is also apparent from the plots in Figure \ref{fig:S}d and f.
Passing from the global minimum to the icosahedral funnel, the
energy first increases as the primary funnel is ascended and then decreases 
during the descent into the second funnel.

From the relative numbers of minima associated with each funnel
it is clear that the icosahedral funnel is much wider.
This is one of the reasons why relaxation is much more
likely to lead to the icosahedral minima.
There are also thermodynamic effects: 
the region containing the icosahedral minima becomes lowest in free energy at low temperature 
($T$$\approx$$0.12 \epsilon k^{-1}$ for LJ$_{38}$\cite{Doye99c,Doye98e} and 
$T$$\approx$$0.09 \epsilon k^{-1}$ for LJ$_{75}$\cite{WalesD97}) because of
the entropy that arises from the large number of low-energy icosahedral minima.
Between this transition temperature and the melting point there is therefore
a thermodynamic driving force towards the icosahedral structures.
Moreover, for LJ$_{38}$ we have shown that the free energy barrier for entering
the icosahedral region of configuration space is lower 
than for entering the fcc funnel.\cite{Doye99c}

Once the cluster enters the icosahedral funnel it is likely to be trapped 
because of the large energy and free energy barriers 
to passing between the two funnels. The energy barriers are 
$4.22\epsilon$ and $3.54\epsilon$ for LJ$_{38}$ and $8.69\epsilon$ and $7.48\epsilon$ for LJ$_{75}$.
At higher temperatures the entropy of the intermediate states reduces
the free energy barriers from these zero temperature limits.\cite{Doye99c} 
For the above reasons LJ$_{38}$ and LJ$_{75}$ 
provide relatively difficult test cases for any unbiased global optimization algorithm. 
It is only relatively recently that algorithms have begun to find the truncated 
octahedron,\cite{Pillardy,Leary97,WalesD97,Wolf98,Deaven96,Niesse96a,Barron96,White98,Pappu98,Michaelian98}
and only one unbiased method\cite{WalesD97} and one method that involves seeding\cite{Wolf98}
have reported finding the Marks decahedron.

\begin{center}
\begin{figure}
\epsfig{figure=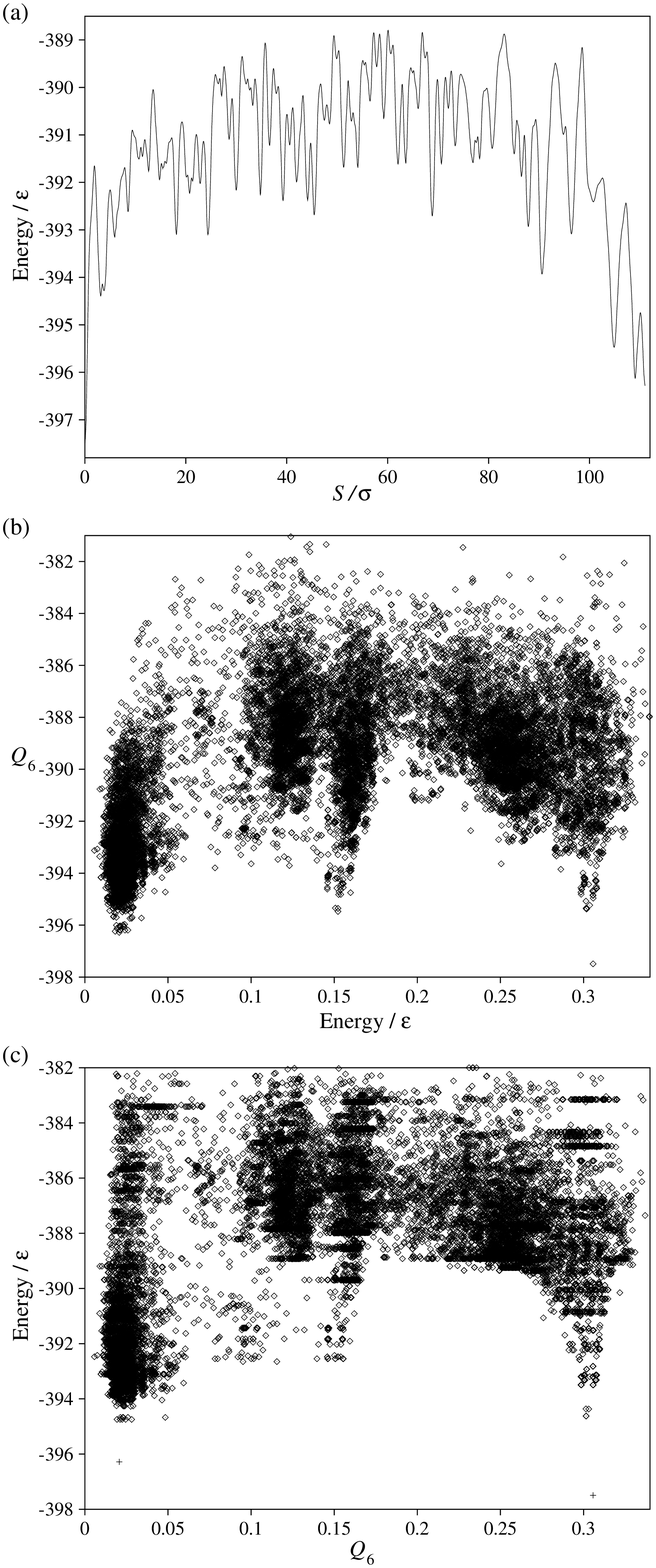,width=8.0cm}
\vglue0.1cm
\begin{minipage}{8.5cm}
\caption{\label{fig:75} Properties of the LJ$_{75}$ PES.
(a) The lowest-barrier path between the global minimum and 
the second lowest minimum. Each maximum corresponds to a transition state.
(b) Scatter plot of the bond order parameter, $Q_6$, versus potential energy for all the
minima in our sample.
(c) Scatter plot of $Q_6$ for each minimum
versus the energy of the highest transition state on the lowest-energy path 
to either structure 75.1 or 75.2, whichever path is the lower.
The two + symbols indicate the values of $Q_6$ and potential energy for the two lowest-energy minima.
}
\end{minipage}
\end{figure}
\end{center}

The difficulty in finding the global minimum for these clusters is illustrated by
statistics for the basin-hopping algorithm.\cite{WalesS99} 
The time required to find the global minimum
for LJ$_{38}$ is a maximum with respect to neighbouring sizes and for LJ$_{75-77}$
the time is so long that good statistics for the first 
passage time have not yet been obtained.
Global optimization is an order of magnitude more difficult 
for LJ$_{75}$ than LJ$_{38}$ for a combination of reasons.
First, LJ$_{75}$ has a much larger search space and a much greater number of minima than LJ$_{38}$.
Second, the temperature at which the global free energy minimum becomes associated with the 
global potential energy minimum lies further below the melting point
($T_{\rm m}$$\approx$$ 0.17 \epsilon k^{-1}$ for LJ$_{38}$\cite{Doye98e} and 
$T_{\rm m}$$\approx$$ 0.29 \epsilon k^{-1}$ for LJ$_{75}$).
Third, the path between the global minimum and the lowest-energy icosahedral structure
is also longer, more complicated and higher in energy than for LJ$_{38}$.
The transition involves not only a change in morphology, but also a change in shape---the Marks decahedron
is oblate whereas the icosahedral structures are prolate. The pathway therefore involves both 
cooperative rearrangements, where the structure twists around 
a quasi-fivefold axis, and surface diffusion steps.
The lowest-energy pathway that we found between the two lowest-energy minima is depicted 
in Figure \ref{fig:75}a. It passes through sixty-five transition states.
There are significantly shorter paths between the same two minima
(the shortest is $41.5\sigma$), but these all involve higher barriers.
There is also a significant difference in the character of the pathways between the
minima at the bottom of the two funnels for LJ$_{38}$ and LJ$_{75}$. 
All the structures along the LJ$_{75}$ pathway are ordered and solid-like, whereas 
at its highest points the LJ$_{38}$ pathway passes through disordered structures.\cite{MarkPhD,Doye97a}

For LJ$_{75}$ the highest-energy minimum on the lowest-energy interfunnel path lies at position $10\,909$
in terms of an energy ranking of the minima in our sample, where 1 is the global minimum.
Many more low-lying minima could have been found
if we had searched the icosahedral region of configuration space more intensively. 
Therefore, if we had simply performed consecutive transition state searches from minima 
in order of their energetic rank it is unlikely that a pathway between the 
two funnels could have been found---rather the search 
would have most likely been stuck in one of the funnels 
cataloguing the multitude of ordered structures with that morphology.
To find a pathway connecting the two lowest-energy minima we therefore had to 
bias the search to probe intermediate structures. 
We first constructed a series of decahedral minima which were decreasingly prolate and
then increasingly oblate.
We then started searching the PES around these structures with the aim of 
finding pathways connecting them both to the Marks decahedron and to the low
energy icosahedral structures.

Once we had found a pathway between the two structures, we only performed
transition state searches from those minima that were connected to either of the two 
lowest-energy minima by a pathway that had no transition states higher than the 
highest-energy transition state on the lowest-energy path between structures 75.1 and 75.2. 
Moreover, this search concentrated on those minima in the set that had an intermediate value of 
the bond-order parameter, $Q_6$.\cite{Steinhardt83,vanD92} 
By this procedure, increasingly low-energy paths were found between the two funnels.
There is, of course, no guarantee that we have found the lowest-energy pathway, 
but we doubt if a significantly lower one exists.

From Figure \ref{fig:75}b we can see how the bond-order parameter can separate
the minima into various groups. The group with $Q_6$$\approx$0.02 correspond
to icosahedral minima with an anti-Mackay overlayer (such as 75.2), 
whereas $Q_6$=0.306 for the Marks decahedron.
The decahedral minima based upon 75.1 have similar values, these being generally lower
for structures that are less oblate. 
There are also bands of minima at intermediate values of $Q_6$. 
For example, icosahedral structures with a Mackay overlayer have $Q_6$$\approx$0.15. 
Some of the other minima with intermediate values have in part motifs
similar to 55.13, and are connected to icosahedral or decahedral minima by rearrangements
involving concerted twists around quasi-fivefold axes.

This distribution of $Q_6$ values provides us with another way of visualizing the
double-funnel structure of the LJ$_{75}$ PES. Figure \ref{fig:75}c maps
out the energy of the transition state at the top of the lowest-barrier 
pathway from each minimum to either 75.1 or 75.2, whichever 
is lower, as a function of $Q_6$ for the minimum.
The plot shows that the barrier separating a decahedral minimum from the Marks 
decahedron increases as the value of $Q_6$ deviates further from 
the value for the global minimum. This trend continues until $Q_6$$\sim$0.16.
Below this value some of the minima have lower barriers for paths to minimum 75.2,
and the energies generally decrease as the value of $Q_6$ approaches that of 75.2
from above. The icosahedral minima with a Mackay overlayer stand out in the plot
as they have both an intermediate value of $Q_6$ and low barriers
connecting them to minimum 75.2. 
The transformations between the two types of icosahedral overlayer are usually
achieved by a concerted twisting of the overlayer around one of the vertices of
the underlying icosahedron.

\subsection{Coarse-graining the PES}
\label{sect:mono}

As the size of the cluster increases our disconnectivity 
graphs focus on an increasingly small proportion of the whole PES
to avoid being swamped by the rapidly increasing number of minima. 
However, it would be desirable to retain a more global picture of the PES. 
To do so, the disconnectivity graphs would need to be
based not on the barriers between minima, but between larger topographical units.
For example, Levy and Becker `diluted' their sample of hexapeptide conformations by
removing conformations that were energetically and structurally similar.\cite{Levy98a}
However, this approach requires a meaningful measure of similarity, which is probably harder to
devise for a cluster than a molecule with a bonded framework. 

Here, we explore the use of monotonic sequences, a concept introduced by Berry
and coworkers,\cite{Kunz95,BerryK95,Ball96,Berry97} to produce a more 
coarse-grained picture of the PES. 
Monotonic sequences are series of connected minima 
where the potential energy decreases with every step. 
The collection of sequences leading to a particular minimum defines a `basin'. 
To avoid confusion with the various other `basins' that have been defined,
we will always refer to such a set as a monotonic
sequence basin (MSB). The MSB leading to the global minimum is termed `primary',
and is separated from neighbouring `secondary' MSB's by transition states
lying on a `primary divide', and so on. It is important to realise that, above
such a divide, it is possible for a minimum to belong to more than one MSB
through different monotonic sequences.

For the division of the PES provided by a monotonic sequence analysis to be useful in an 
analysis of the dynamics, transitions between minima in an MSB must be more rapid than
transitions between different MSB's. Kunz and Berry found evidence for this in a 
simplified model of LJ$_{19}$.\cite{Kunz95,BerryK95} However, there is no 
guarantee that this separation will always hold, because MSB's are defined by the connections between 
minima without reference to the size of the barriers between them.

\begin{center}
\begin{figure}
\epsfig{figure=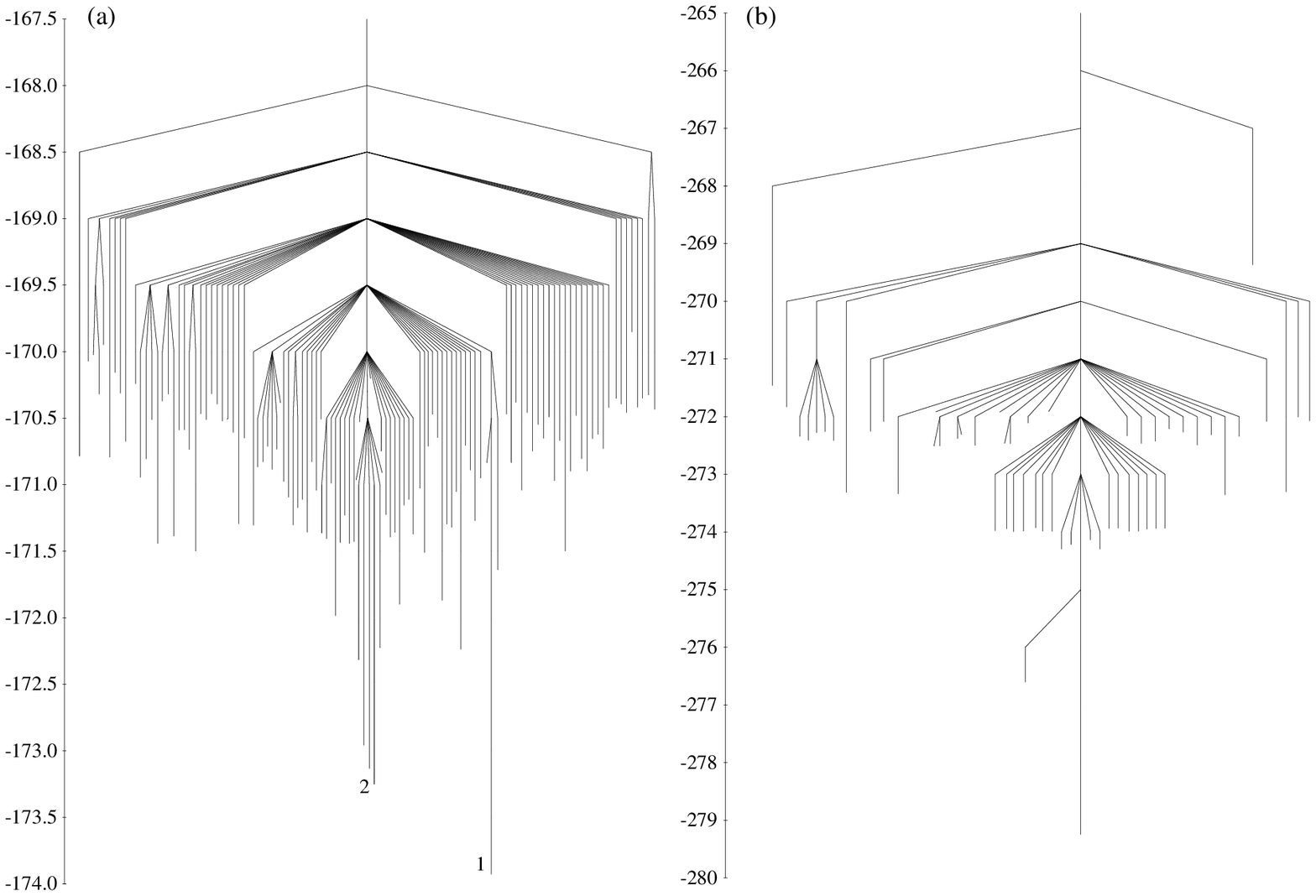,width=16.5cm}
\vglue0.1cm
\begin{minipage}{17.6cm}
\caption{\label{fig:tree.mono} Disconnectivity graphs for
(a) LJ$_{38}$ and (b) LJ$_{55}$. 
Only the branches corresponding to monotonic sequence basin bottoms are shown.
The branches associated with the minima depicted in Figure \ref{fig:structures}
are labelled by their energetic rank.
The energy scale is in units of $\epsilon$.
}
\end{minipage}
\end{figure}
\end{center}

Disconnectivity graphs that only include the minima at the bottom of each MSB 
can be produced by excluding all the minima directly connected to a lower-energy minimum.
In the resulting graphs, branches are joined by a node
at the energy of the lowest transition state on the divide between the MSB's.
For the clusters we consider here, the number of MSB's is small
enough (Table \ref{table:nsp}) for all of them to be represented on the graph. 
Two examples are shown in Figure \ref{fig:tree.mono}.

\end{multicols}

\begin{multicols}{2}

One implication of the method we use to generate our samples of stationary points
is that higher-energy minima that were not used as starting points for
transition state searches are unlikely to lie at the bottom of an MSB. These minima 
lie at the higher end of a pathway that was found in a search from 
a lower-energy minimum. Therefore, most of these minima are directly connected to 
a lower-energy minimum, and so cannot be at the bottom of an MSB. 
Occasionally, the pathways do not connect back to the starting minimum but 
to another unsearched minimum in the sample. 
Only in these rare instances can an unsearched minimum be
at the bottom of an MSB in our sample. Therefore, although the disconnectivity graphs based upon
MSB's provide a more global picture of the PES than the graphs 
in the previous section, 
they are still limited by the incompleteness of our samples. The graphs only provide
a reliable picture of the PES around the $n_{\rm search}$ lowest-energy minima, where 
$n_{\rm search}$ is the number of minima from which transition state searches have been performed.

For LJ$_{13}$ there is only one MSB, reflecting its ideal funnel character and the remarkable 
degree of connectivity.
For  LJ$_{19}$ there are only a few MSB's, and these are all directly connected
to the primary MSB, again reflecting the single funnel character of the PES that we noted earlier.
The double-funnel character of the LJ$_{38}$ PES is still apparent in the MSB disconnectivity graph,
but now we get a better impression of the overall shape of the PES (Figure \ref{fig:tree.mono}a). 
There is a wide, gently sloping funnel down from the higher-energy minima towards the low-energy 
icosahedral funnel, whereas the funnel down to the global minimum is much narrower. 
2292 of the minima lie on monotonic sequences to the lowest-energy icosahedral minimum,
whereas only 518 lie on sequences leading to the global minimum. Only 12 of the minima lie on sequences
to both, showing that there is little overlap between the two MSB's.

For LJ$_{55}$ the MSB analysis leads to a remarkable simplification of the disconnectivity
graph (Figure \ref{fig:tree.mono}b). The single-defect minima produce just one MSB, and the fine
structure of the two-defect minima collapses onto the band of MSB's that branch off at $-273\epsilon$
and $-272\epsilon$. The remaining branches are mostly three-defect minima with the three surface
atoms close together. The graph clearly shows the single-funnel character of the PES.

\end{multicols}

\begin{center}
\begin{figure}
\epsfig{figure=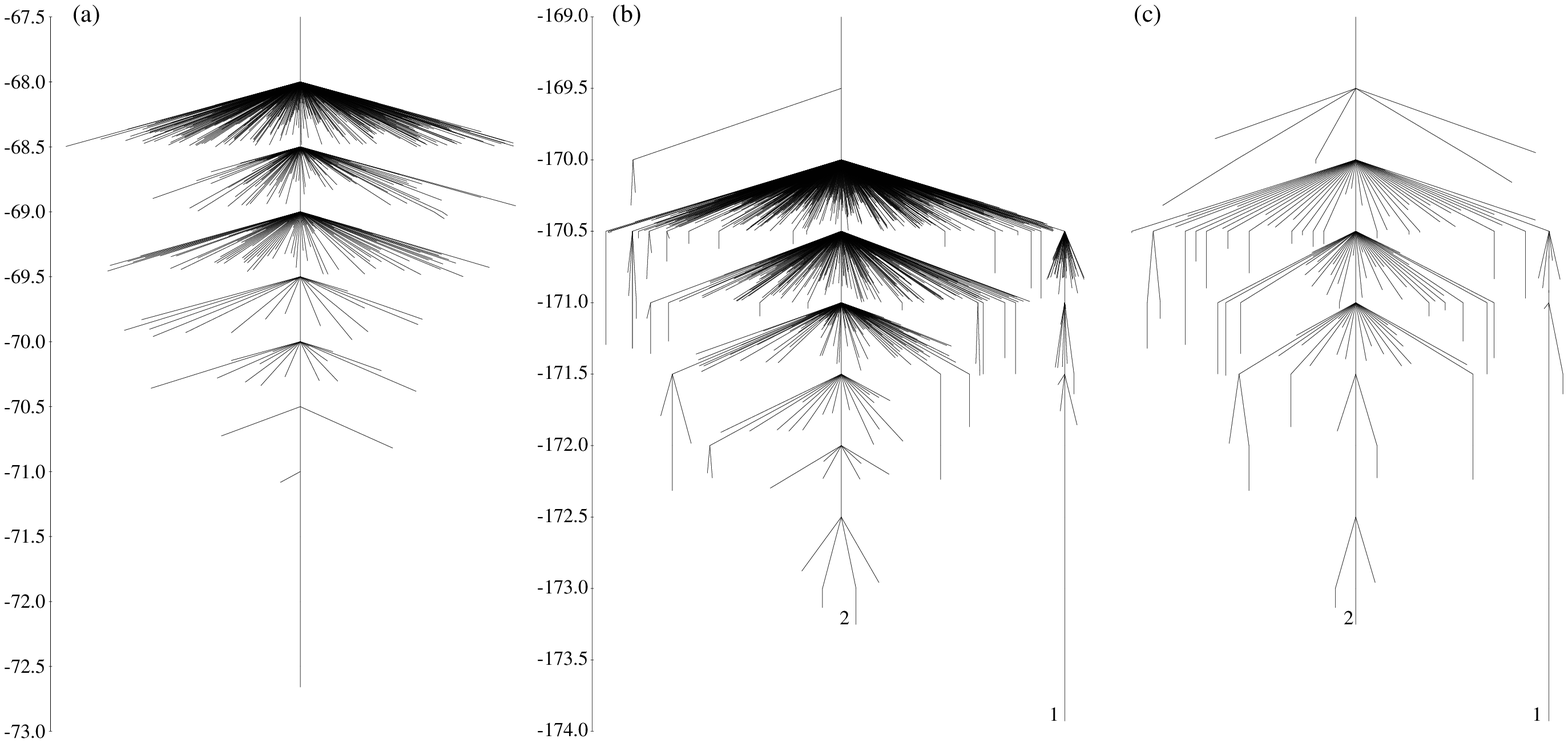,width=16.5cm}
\vglue0.1cm
\begin{minipage}{17.6cm}
\caption{\label{fig:tree.nobar} Disconnectivity graphs for the transformed energy
surfaces of (a) LJ$_{19}$ and (b) and (c) LJ$_{38}$. 
In (a) and (b) only the branches corresponding to the lowest $n_{\rm search}$ steps are shown.
In (c) only the branches corresponding to monotonic sequence basin bottoms are shown.
The branches associated with the minima depicted in Figure \ref{fig:structures}
are labelled by their energetic rank. 
The energy scale is in units of $\epsilon$.
}
\end{minipage}
\end{figure}
\end{center}

\begin{multicols}{2}

\subsection{Transforming the PES}

\label{sect:gmin}

\begin{figure}
\begin{center}
\epsfig{figure=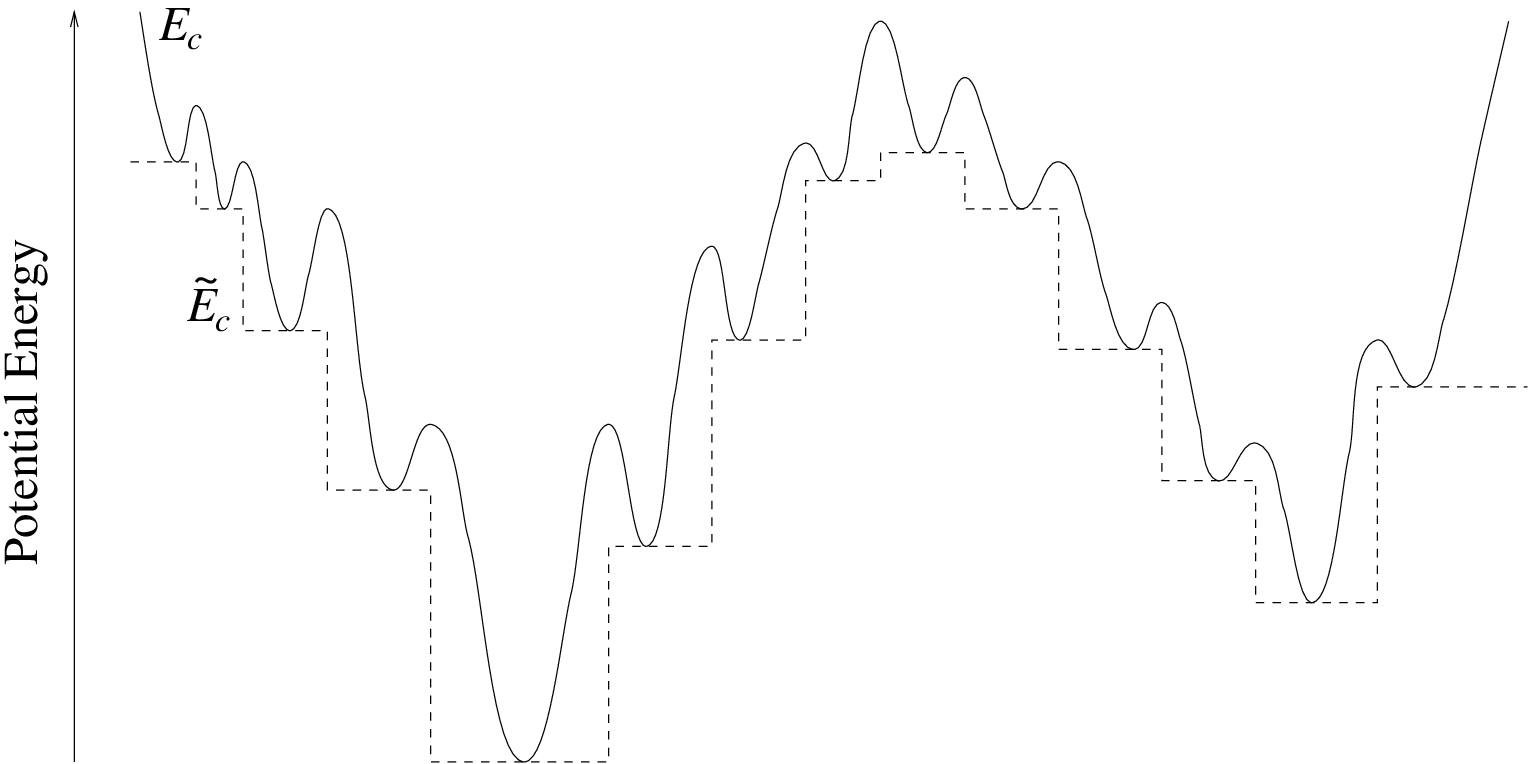,width=8.2cm}
\vglue 0.2cm
\begin{minipage}{8.5cm}
\caption{\label{fig:trans}
A schematic diagram illustrating the effects of our potential
energy transformation for a one-dimensional example.
The solid line is the potential energy of the original surface and the dashed line is the
transformed energy, $\tilde E_c$.}
\end{minipage}
\end{center}
\end{figure}

One approach to global optimization is to transform the PES to a form 
for which it is easier to find the global minimum.\cite{StillW88} One efficient 
approach of this type is the Monte Carlo minimization\cite{Li87a} or 
basin-hopping\cite{WalesD97} algorithm. This unbiased method succeeded in finding
the global minimum for all the clusters considered in this paper.\cite{WalesD97}
In this approach the transformed potential energy $\tilde E_c$ is given by 
\begin{equation}
\tilde E_c({\bf X}) = {\rm min}\left\{ E_c({\bf X}) \right\}.
\end{equation}
Hence, the potential energy at any point in configuration space is assigned to that of the local
minimum obtained by a minimization starting from that point, and the PES is mapped
onto a set of interpenetrating staircases with steps corresponding to the 
basins of attraction surrounding a particular local minimum. 
Figure \ref{fig:trans} illustrates the transformation for a simple one-dimensional example.
Note that the transformation does not change the identity of the global minimum, 
nor the relative energies of any of the minima, 
but it does remove the downhill barriers between directly connected minima.
This latter change has a significant effect on the dynamics and 
thermodynamics.\cite{Doye98e,Doye98a}

Due to our PES search strategy virtually all the unsearched 
higher-energy minima are directly connected to lower-energy minima. As the transformation 
removes the downhill barriers, in the disconnectivity graphs these unsearched minima would be directly 
connected to the stem associated with the superbasin which contains 
the relevant lower-energy minimum. 
This connectivity makes the higher-energy regions of the PES look funnel-like 
irrespective of their actual character. 
Therefore, in the graphs we only depict branches which lead to the 
lowest $n_{\rm search}$ steps. 

The two examples shown in Figure \ref{fig:tree.nobar} illustrate the 
effects of the staircase transformation on the disconnectivity graphs.
It removes the barriers to progress down a funnel, and 
the disconnectivity graph for LJ$_{19}$ is therefore transformed into that for an ideal funnel.
The long, dangling branches that are indicative of large barriers 
have disappeared, and so relaxation is now easier.
However, barriers to interfunnel passage remain because the 
major component of such barriers usually arises from the high energy minima 
that the system has to pass through to go between funnels. (Figure \ref{fig:75}a). 
For example, the transformation reduces the potential energy barriers 
for interfunnel passage by only $0.68\epsilon$ for LJ$_{38}$ 
and $0.86\epsilon$ for LJ$_{75}$. 
Therefore, the splitting of the LJ$_{38}$ disconnectivity graph into two funnels is still clear, 
and perhaps even more obvious because many of the other barriers have been 
removed (Figure \ref{fig:tree.nobar}b). 
As for LJ$_{19}$, the icosahedral region of configuration space now looks much 
more funnel-like.

The retention of the energy barriers for interfunnel passage in LJ$_{38}$
and LJ$_{75}$ means that the global minima of these clusters are still hard to find
even on the transformed PES.\cite{WalesS99} 
In fact, the success of the basin-hopping algorithm for these clusters lies in the changes to the
thermodynamics caused by the transformation.\cite{Doye98e,Doye98a} 
The thermodynamic transitions are broadened, thus providing a temperature window 
where both the states at the top of the paths for interfunnel passage and the states at the bottom of 
the two funnels have a significant probability of being occupied,
making passage between funnels easier.

On the transformed PES any pathway that is monotonically decreasing in energy
(albeit a stepped rather than a smooth one)
must end at the step corresponding to the bottom of an MSB. 
Therefore, an MSB on the transformed PES is analogous 
to a basin of attraction surrounding a minimum on the original PES. 
Figure \ref{fig:trans} illustrates this point: 
on the transformed PES the two MSB's are like `stepped' minima.
However, the analogy breaks down in one respect.
A basin of attraction surrounding a minimum is defined as the set of points from 
which steepest-descent paths lead down to that minimum.
The use of the steepest-descent path ensures that each point in
configuration space is mapped uniquely to a minimum even when a point is higher
in energy than the lowest transition state connected to that minimum.
However, on a step on the transformed PES the gradient is zero 
and so no steepest-descent direction is defined.  
Therefore, the mapping of a point in configuration space
to a MSB bottom may not be unique if it lies above the divide between two MSB's,
and so MSB's, unlike basins of attraction, can overlap.

In Figure \ref{fig:tree.nobar}c we show the disconnectivity graph for the transformed PES 
of LJ$_{38}$ with only the branches leading to the bottoms of MSB's displayed. It has 
a similar structure to Figure \ref{fig:tree.nobar}b but the density of branches 
is greatly reduced.

\section{Conclusions}

In this paper we have shown that disconnectivity graphs can provide a 
valuable tool for visual representation of a PES, and that they can 
provide clear physical insights into structure, dynamics and thermodynamics.

The exponential increase in the number of minima with system size
means that disconnectivity graphs with branches representing individual minima 
must, as the size increases, inevitably focus on low-energy regions of the PES 
which represent a decreasing proportion of the whole configuration space.
By taking monotonic sequences basins as the basic topographical unit, we
have extended the cluster size for which a disconnectivity graph can represent
the full sample of minima. However, these graphs still cannot provide
a truly global picture of the PES because of the incompleteness of our samples of 
stationary points.

We have also used disconnectivity graphs to probe the transformed PES
that is searched by the Monte Carlo minimization or `basin-hopping' global 
optimization algorithm. 
The transformation removes downhill barriers, making relaxation down 
a funnel much easier. 
For example, the algorithm can find the global minimum of LJ$_{55}$, which 
has a single-funnel PES, in on average fewer than 150 steps when 
started from a random geometry.\cite{Doye98e,WalesS99}
However, the barriers between funnels remain, and so 
the success of basin-hopping for double-funnel PES's
must be explained in terms of the different thermodynamics
for the transformed landscape.\cite{Doye98e,Doye98a} 

\acknowledgements

J.P.K.D. is the Sir Alan Wilson Research Fellow at Emmanuel College, Cambridge.
D.J.W.\ is grateful to the Royal Society 
and M.A.M.\ to the Engineering and Physical Sciences Research Council 
for financial support. We would also like to thank Dr David Manolopoulos
for his suggestion that disconnectivity graphs could be used to probe
the transformed energy landscape used by the basin-hopping algorithm.

\end{multicols}

\begin{thebibliography}{10}

\bibitem{StillW88}
F.~H. Stillinger and T.~A. Weber, J. Stat. Phys. {\bf 52},  1429  (1988).

\bibitem{Angell95}
C.~A. Angell, Science {\bf 267},  1924  (1995).

\bibitem{StillW84a}
F.~H. Stillinger and T.~A. Weber, Science {\bf 225},  983  (1984).

\bibitem{Wales93a}
D.~J. Wales, Mol. Phys. {\bf 78},  151  (1993).

\bibitem{Doye95a}
J.~P.~K. Doye and D.~J. Wales, J. Chem. Phys. {\bf 102},  9659  (1995).

\bibitem{Czerminski90}
R. Czerminski and R. Elber, J. Chem. Phys. {\bf 92},  5580  (1990).

\bibitem{Kunz95}
R.~E. Kunz and R.~S. Berry, J. Chem. Phys. {\bf 103},  1904  (1995).

\bibitem{Angelani98}
L. Angelani, G. Parisi, G. Ruocco, and G. Viliani, Phys. Rev. Lett. {\bf 81},
  4648  (1998).

\bibitem{MarkPhD}
M.~A.~Miller, PhD Thesis, {\it Energy Landscapes and Dynamics of Model
  Clusters}, University of Cambridge (1999).

\bibitem{Becker97}
O.~M. Becker and M. Karplus, J. Chem. Phys. {\bf 106},  1495  (1997).

\bibitem{Levy98a}
Y. Levy and O.~M. Becker, Phys. Rev. Lett. {\bf 81},  1126  (1998).

\bibitem{WalesMW98}
D.~J. Wales, M.~A. Miller, and T.~R. Walsh, Nature {\bf 394},  758  (1998).

\bibitem{Miller99a}
M.~A. Miller, J.~P.~K. Doye, and D.~J. Wales, J. Chem. Phys. {\bf 110},  328
  (1999).

\bibitem{Doye99c}
J.~P.~K. Doye, M.~A. Miller, and D.~J. Wales, J. Chem. Phys.  in press
  (cond-mat/9808265).

\bibitem{Pillardy}
J. Pillardy and L. Piela, J. Phys. Chem. {\bf 99},  11805  (1995).

\bibitem{Leary97}
R.~H. Leary, J. Global Optimization {\bf 11},  35  (1997).

\bibitem{WalesD97}
D.~J. Wales and J.~P.~K. Doye, J. Phys. Chem. A {\bf 101},  5111  (1997).

\bibitem{HoareM76}
M.~R. Hoare and J. McInnes, Faraday Discuss., Chem. Soc. {\bf 61},  12  (1976).

\bibitem{Tsai93a}
C.~J. Tsai and K.~D. Jordan, J. Phys. Chem. {\bf 97},  11227  (1993).

\bibitem{Still99}
F.~H. Stillinger, Phys. Rev. E {\bf 59},  48  (1999).

\bibitem{BerryK95}
R.~S. Berry and R.~E. Breitengraser-Kunz, Phys. Rev. Lett. {\bf 74},  3951
  (1995).

\bibitem{Li87a}
Z. Li and H.~A. Scheraga, Proc. Natl. Acad. Sci. USA {\bf 84},  6611  (1987).

\bibitem{LJ}
J.~E. Jones and A.~E. Ingham, Proc. R. Soc. A {\bf 107},  636  (1925).

\bibitem{Pancir74a}
J. Panc{\'\i}\v{r}, Coll. Czech. Chem. Comm. {\bf 40},  1112  (1974).

\bibitem{Cerjan}
C.~J. Cerjan and W.~H. Miller, J. Chem. Phys. {\bf 75},  2800  (1981).

\bibitem{Wales94b}
D.~J. Wales, J. Chem. Phys. {\bf 101},  3750  (1994).

\bibitem{Page88}
M. Page and J.~W. McIver, J. Chem. Phys. {\bf 88},  922  (1988).

\bibitem{Mackay}
A.~L. Mackay, Acta Cryst. {\bf 15},  916  (1962).

\bibitem{Raoult89a}
B. Raoult, J. Farges, M.~F. de~Feraudy, and G. Torchet, Phil. Mag. B {\bf 60},
  881  (1989).

\bibitem{Leopold}
P.~E. Leopold, M. Montal, and J.~N. Onuchic, Proc. Natl. Acad. Sci. USA {\bf
  89},  8271  (1992).

\bibitem{Bryngelson95}
J.~D. Bryngelson, J.~N. Onuchic, N.~D. Socci, and P.~G. Wolynes, Proteins:
  Structure, Function and Genetics {\bf 21},  167  (1995).

\bibitem{Doye95c}
J.~P.~K. Doye, D.~J. Wales, and R.~S. Berry, J. Chem. Phys. {\bf 103},  4234
  (1995).

\bibitem{Doye98e}
J.~P.~K. Doye, D.~J. Wales, and M.~A. Miller, J. Chem. Phys. {\bf 109},  8143
  (1998).

\bibitem{sdvef}
Some of these values are slightly different from those mentioned in Ref.\
  \onlinecite{WalesD97b} because the pathways were calculated here using
  steepest-descent rather than eigenvector-following.

\bibitem{Doye95b}
J.~P.~K. Doye and D.~J. Wales, J. Chem. Phys. {\bf 102},  9673  (1995).

\bibitem{Kunz93}
R.~E. Kunz and R.~S. Berry, Phys. Rev. Lett. {\bf 71},  3987  (1993).

\bibitem{Kunz94}
R.~E. Kunz and R.~S. Berry, Phys. Rev. E {\bf 49},  1895  (1994).

\bibitem{Wolf98}
M.~D. Wolf and U. Landman, J. Phys. Chem. A {\bf 102},  6129  (1998).

\bibitem{Wales93c}
D.~J. Wales, J. Chem. Soc., Faraday Trans. {\bf 89},  1305  (1993).

\bibitem{Ball96}
K.~D. Ball, R.~S. Berry, R.~E. Kunz, F.-Y. Li, A. Proykova, and D.~J. Wales, 
Science {\bf 271},  963  (1996).

\bibitem{Echt81}
O. Echt, K. Sattler, and E. Recknagel, Phys. Rev. Lett. {\bf 47},  1121
  (1981).

\bibitem{Northby87}
J.~A. Northby, J. Chem. Phys. {\bf 87},  6166  (1987).

\bibitem{WalesS99}
D.~J. Wales and H.~A. Scheraga, Science, submitted  (1999).

\bibitem{38again}
Although the disconnectivity graph for LJ$_{38}$ has been presented
  previously,\cite{WalesMW98,Doye99c} we present it again here to facilitate
  the comparisons with the disconnectivity graphs of the other clusters and
  with the alternative forms of the graph given in Sections \ref{sect:mono} and
  \ref{sect:gmin}. For a detailed discussion of the fine structure of the
  LJ$_{38}$ graphs see Ref.\ \onlinecite{Doye99c}.

\bibitem{Deaven96}
D.~M. Deaven, N. Tit, J.~R. Morris, and K.~M. Ho, Chem. Phys. Lett. {\bf 256},
  195  (1996).

\bibitem{Niesse96a}
J.~A. Niesse and H.~R. Mayne, J. Chem. Phys. {\bf 105},  4700  (1996).

\bibitem{Barron96}
C. Barr\'on, S. G\'omez, and D. Romero, Appl. Math. Lett. {\bf 9},  75  (1996).

\bibitem{White98}
R.~P. White, J.~A. Niesse, and H.~R. Mayne, J. Chem. Phys. {\bf 108},  2208
  (1998).

\bibitem{Pappu98}
R.~V. Pappu, R.~K. Hart, and J.~W. Ponder, J. Phys. Chem. B {\bf 102},  9725
  (1998).

\bibitem{Michaelian98}
K. Michaelian, Chem. Phys. Lett. {\bf 293},  202  (1998).

\bibitem{Doye97a}
J.~P.~K. Doye and D.~J. Wales, Z. Phys. D {\bf 40},  194  (1997).

\bibitem{Steinhardt83}
P.~J. Steinhardt, D.~R. Nelson, and M. Ronchetti, Phys. Rev. B {\bf 28},  784
  (1983).

\bibitem{vanD92}
J.~S. van Duijneveldt and D. Frenkel, J. Chem. Phys. {\bf 96},  4655  (1992).

\bibitem{Berry97}
R.~S. Berry, N. Elmaci, J.~P. Rose, and B. Vekhter, Proc. Natl Acad. Sci. USA
  {\bf 94},  9520  (1997).

\bibitem{Doye98a}
J.~P.~K. Doye and D.~J. Wales, Phys. Rev. Lett. {\bf 80},  1357  (1998).

\bibitem{WalesD97b}
D.~J. Wales and J.~P.~K. Doye, J. Chem. Phys. {\bf 106},  5296  (1997).

\end{thebibliography}
\end{document}